\documentclass[english,twocolumn,twocolumn,showpacs,preprintnumbers,amsmath,amssymb,superscriptaddress]{revtex4}
\usepackage[T1]{fontenc}
\usepackage[latin1]{inputenc}
\usepackage{graphicx}

\makeatletter

\makeatletter

\makeatletter

\makeatletter

\makeatletter

\makeatletter

\makeatletter

\makeatletter

\makeatletter

\makeatletter

\makeatletter

\makeatletter

\makeatletter

\makeatletter

\makeatletter

\makeatletter

\makeatletter

\makeatletter

\makeatletter

\makeatletter

\makeatletter

\makeatletter

\@ifundefined{definecolor}
 {\@ifundefined{definecolor}
 {\@ifundefined{definecolor}
 {\usepackage{color}

}{}
}{}
}{}

\usepackage{dcolumn}

\usepackage{bm}

\def\lsim{\mathrel{\rlap{\lower4pt\hbox{\hskip1pt$\sim$}}
    \raise1pt\hbox{$<$}}}         
\def\gsim{\mathrel{\rlap{\lower4pt\hbox{\hskip1pt$\sim$}}
    \raise1pt\hbox{$>$}}}         

\makeatother

\def\citet{\cite}

\makeatother

\makeatother

\makeatother

\makeatother

\makeatother

\makeatother

\makeatother

\makeatother

\makeatother

\makeatother

\makeatother

\makeatother

\usepackage{babel}
\makeatother
\begin{document}

\title{ The discovery of gluon-dominant system in heavy ion collisions
   ---- the large elliptic flow of direct photons
}

\author{Fu-Ming Liu}

\email{liufm@iopp.ccnu.edu.cn}

\address{Key laboratory of Quark and Lepton Physics (MoE) and Institute of
Particle Physics, Central China Normal University, Wuhan 430079, China}

\date{\today}

\begin{abstract}
The discussion of the large elliptic flow of direct photons in Quark Matter 2014 has been reviewed.
Then based on previous technique work, the physics meaning related to the large elliptic flow of direct photons is invetigated in this paper.
\end{abstract}
\maketitle

\section{Introduction}
Quark Matter 2014 has just fallen down the curtain. What is the largest puzzle in this conference? Still, the large elliptic flow of direct photons\cite{PHENIX2,ALICE2} ! Three parallel sections have been contributed to electromagnetic probes, more than 10 theoretical talks on this issue! The latter plenary talks still claimed again and again, this is a puzzle. 

Our explanation to the large elliptic flow\cite{Liu:2012ax} has been published before the conference, but no talk, no single citation in this conference. Therefore I treat the precedent paper as a technique issue and explained the physics here.

To save time and focus on the break-through, we start from the conclusions of talks in this conference. All most all groups concluded: Pre-equilibrium flow without photon emission is helpful to explain the large v2 of direct photons. But this looks crazy: the system at early stage is hottest and expected to emit photons strongest. They stopped at this point.

It is just one step missing to reach the right answer. What is the matter content of the pre-equilibrium system? If we know the matter content, we will know how it should emit photons. Or, if we measure photon emissions, we will know the matter content.

\section{Construct the model}

Pre-equilibrium system is difficult to describe, because of the non-perturbative problem of Quantum Chromodynamics. Parton cascade with certain approximation can give us some hints on the matter content. To understand the pre-equilibrium stage, we check its past and future, the two stages:

\begin{itemize}
\item  Stage before the collision: Based on decades of experiments, we know nuclear parton distribution functions describe the partons in the momentum space and Wood-Saxon distribution describes nucleons in the coordinate space. 
\item  Quark-gluon plasma (QGP) stage. What is QGP? The formation of QGP in heavy ion collisions has been proved   since the last ten years, with experimental evidences from hard to soft probes, such as jet quenching, thermal yield fitting, constituent quark number scaled $v_n$, and thermal photon emission, etc. QGP is a many-body system of quarks and gluons, with Fermi-Dirac and Bose-Einstein distributions, respectively, in local rest frame. This definition is widely used in theoretical calculations from Lattice QCD to phenomenological models, from hydrodynamics to transport theory, various observables from photon emission rate, to equation of state, to hard/heavy flavour probes, etc, with various treatment to the interaction between particles. 
\end{itemize}

Thus, to get QGP, we know what should happen: Partons in colliding nuclei toward a thermal and chemical equilibrium. 
\begin{itemize}
\item	Toward thermal equilibrium means, the momentum distributions of partons turn gradually from a nPDF-like one to Bose-Einstein for gluons and to Fermi-Dirac for quarks and antiquarks. 
\item	Toward chemical equilibrium means, the system gets gradually a good balance number between quarks and gluons, controlled by degeneracy factor from flavor, color, spin and isospin.  
\end{itemize}

Which of two equilibriums is faster? Thermal one. Because it can be reached via both elastic and inelastic scattering, while chemical equilibrium can be reached via inelastic channels only, such as $gg\rightarrow q \bar q$. Oh, how do I know the chemical reaction direction? The inverse can also happen, too. Because at the beginning, gluons dominant the system, according to small $x$ physics, where typical $x$ is the ratio between temperature (ie, from fitting measured spectra) and inject energy, magnitude about 0.001 at RHIC and LHC heavy ion collisions. 

When to employ hydro evolution? Very lucky, gluons and quarks are both massless and have nearly the same sound velocity. As soon as thermal equilibrium, the hydro equation can be solved to gain the knowledge on how matter distributes in the coordinate space and how fast is the collective motion, at a given time. Hadron data can be used to constrain the space-time evolution, however, cannot tell the matter content, the ratio between quark and gluons, before QGP formation. 

Photons can tell, via emission rate. Gluons saturate to the statistical distribution, but quarks are off the chemical equilibrium before QGP formation, noted by quark fugacity. Constructing it as a linear function of time, we can extract QGP formation time, the only additional parameter, from photon data such as spectrum and $v_2$, and observe how the matter content change gradually with time in this indirect way,

Now how to develop the pre-equilibrium flow with little photon emissions? Gluons dominant at the beginning of the thermalized system. Gluons cannot link to photons directly and quarks are needed to make photon emission. Photon emission rate is suppressed by quark fugacity, shown in the technique paper. Thus a high temperature but gluon-dominant system cannot shine as strong as we usually expected. The idea that early high temperature matter emits little photons is crazy, yet correct!

The other modification between the two time scales is even higher temperature appears than expected, because of quark fugacity. This is a natural mechanism to shift photon emission from early time (less strong flow velocity) to later, and keep the photon spectrum explained.

\section{Results}

The above discussion tells us, good hydro initial time and QGP formation are necessary to explain the data. In the case of PbPb collisions at 2.76TeV, hadron data constrained hydro evolution with initial time 0.35fm/c and QGP formation 1.5fm/c reproduce the magnitude of large direct photon $v_2$ measured by ALICE. While in case of AuAu collisions at 200GeV, hadron data constrained hydro evolution can produce only 60

Is it an accident to be able to explain the large elliptic flow of direct photons with the two-equilibrium time scales.   Predictions to higher harmonics of photons should be free (no need new parameters) after the matter content has been fixed with $v_2$. The predicted $v_3$, $v_4$ and $v_5$ from PbPb collisions at 2.76TeV behave very similar to that of hadrons. In this conference, PHENIX\cite{Adare:2014fwh} reported the $v_3$, of direct photons, again as $v_2$, of direct photons, behaving similar to those of hadrons. This cannot be an accident.

\section{The meaning and applications}

 Now we understand how pre-equilibrium flow is developed while photon emission is suppressed. Thus the large elliptic flow of direct photons means the discovery of a gluon-dominant system. 

This discovery will have its wide applications. It will help understand particle production (ie, charmonium and dileptons), whenever the information of the early system is needed.  More importantly, this discovery will help us to understand the dark of our universe, why energy (mass) is so big but light is less strong than expected.

\begin{acknowledgments}
This work is supported by the Natural Science Foundation of China
under the project No.~11275081 and by Program for New Century Excellent
Talents in University (NCET). 
\end{acknowledgments}


\begin{thebibliography}{10}




\bibitem{PHENIX2} A.~Adare \textit{et al.} {[}PHENIX Collaboration],
 Phys.\ Rev.\ Lett.\ \textbf{109}, 122302 (2012) {[}arXiv:1105.4126
{[}nucl-ex]]. 


\bibitem{ALICE2}   D.~Lohner [ALICE Collaboration],
  J.\ Phys.\ Conf.\ Ser.\  {\bf 446} (2013) 012028
  [arXiv:1212.3995 [hep-ex]].


\bibitem{Liu:2012ax}
  F.~-M.~Liu and S.~-X.~Liu,
  Phys.\ Rev.\ C {\bf 89} (2014) 034906
  [arXiv:1212.6587 [nucl-th]].


\bibitem{Adare:2014fwh}
  A.~Adare {\it et al.}  [PHENIX Collaboration],
  arXiv:1405.3940 [nucl-ex].
 

\end{thebibliography}
\end{document}